\documentclass[11pt]{article}

\usepackage{amstext}
\usepackage{enumerate}
\usepackage{citesort}
\usepackage{epsfig}

\newcounter{hgt}

\def\myendproof{{\ \vbox{\hrule\hbox{%
   \vrule height1.3ex\hskip0.8ex\vrule}\hrule }}\par}
\newtheorem{theorem}{Theorem}[section]
\newtheorem{lemma}[theorem]{Lemma}
\newtheorem{corollary}[theorem]{Corollary}
\newenvironment{proof}{{\it Proof. }}{\myendproof}

\newcommand{\floor}[1]{\left\lfloor{#1}\right\rfloor}
\newcommand{\ceiling}[1]{\left\lceil{#1}\right\rceil}
\newcommand{\less}[1]{\phi\left(#1\right)}
\newcommand{\setof}[1]{\left\{{#1}\right\}}
\newcommand{\set}[2]{\left\{{#1}\mid{#2}\right\}}
\newcommand{\adv}{{\cal A}}
\newcommand{\bid}{{\cal D}}
\newcommand{\equilibrium}{\text{\rm equilibrium}}
\newcommand{\unif}{\pi_{\text{\rm unif}}}
\newcommand{\id}{\pi_{\text{\rm id}}}
\newcommand{\msum}[1]{\text{\rm sum}({#1})}
\newcommand{\bsum}[1]{\text{\rm bsum}({#1})}
\newcommand{\ratio}{R}
\newcommand{\budget}{\beta}
\newcommand{\ourbidset}{\Psi}
\newcommand{\winning}[1]{w^*({#1})}
\newtheorem{fact}{Fact}

\title{Optimal Bid Sequences for Multiple-Object Auctions \\
with Unequal Budgets\thanks{A preliminary version to appear in {\em
Proceedings of the 11th Annual International Symposium on Algorithms
and Computation} (ISAAC 2000).}}
\date{\today}
\author{
  Yuyu Chen\thanks{Department of Computer Science, Yale University,
  New Haven, CT 06520, USA (chen-yuyu@cs.yale.edu).  This author's
  research was supported in part by NSF grant CCR-9531028.}
\and 
  Ming-Yang Kao\thanks{Department of Computer Science, Yale
  University, New Haven, CT 06520, USA
  (kao-ming-yang@cs.yale.edu). This author's research was supported in
  part by NSF grants CCR-9531028 and CCR-9988376.}
\and 
  Hsueh-I Lu\thanks{Institute of Information Science, Academia Sinica,
  Taipei 115, Taiwan, R.O.C. (hil@iis.sinica.edu.tw). This author's
  research was supported in part by NSC grant NSC-89-2213-E-001-034.}
}

\begin{document}

\maketitle
\begin{abstract}
In a multiple-object auction, every bidder tries to win as many
objects as possible with a bidding algorithm. This paper studies {\em
position-randomized auctions}, which form a special class of
multiple-object auctions where a bidding algorithm consists of an
initial bid sequence and an algorithm for randomly permuting the
sequence. We are especially concerned with situations where some
bidders know the bidding algorithms of others. For the case of only
two bidders, we give an optimal bidding algorithm for the
disadvantaged bidder. Our result generalizes previous work by allowing
the bidders to have unequal budgets.  One might naturally anticipate
that the optimal expected numbers of objects won by the bidders would
be proportional to their budgets.  Surprisingly, this is not true.
Our new algorithm runs in optimal $O(n)$ time in a straightforward
manner.  The case with more than two bidders is open.
\end{abstract}

%\begin{keywords}
%auction theory, bidding algorithms, electronic commerce, automated
%negotiation mechanisms, software agents, market-based control
%\end{keywords}

%\begin{AMS}
%  05A99, 60C05, 68R05, 90A09, 90A12, 90D10, 90D13
%\end{AMS}

\section{Introduction}
Economists have long recognized the usefulness of auction as a means
of price determination without intermediary market makers. As a
result, there already exists an enormous Economics literature on
auction theory and practice (see, e.g.,
\cite{MW82,Mye81,PS88,HP93,McA87,Wil92}).  Relatively recently,
computer scientists have become aware of the potential efficiency of
auction as a general method of resource allocation~\cite{clearwater}.
For instance, Gagliano, Fraser, and Schaefer~\cite{GFM95} applied
auction techniques to allocating decentralized network
resources. Bertsekas \cite{Bertsekas1995} designed an auction-type
algorithm for the classical maximum flow problem.

With the advent of the Word Wide Web, Internet-based auction is
rapidly becoming an essential buying and selling medium for both
individuals and organizations. It is projected that most of the future
Internet auctions will necessarily be conducted by software agents
instead of human bidders and auctioneers
{\cite{Preist99,HuhnsV99,RMNGS98,WurmanWW98}}.  Consequently, there is
an increasing need for highly efficient and sophisticated auction
mechanisms and bidding algorithms.  To meet this need, Computer
Science is witnessing heightened research efforts on such mechanisms
and algorithms. Among the several basic research themes that have
emerged from these efforts, the following three are particularly
relevant to this paper.

The first theme is multiple-object auction
\cite{DKLP1999,AADKao2000.scp,FLS99,PU2000,LOS1999,RPH1998},
where each bidder may bid on several objects simultaneously instead of
one at a time. The second theme is the informational security of
auction.  For instance, Cachin \cite{cachin1999} and Stajano and
Anderson~\cite{SA99} were concerned with the privacy of bidders.
Sako~\cite{Sako2000} discussed how to hide information about losing
bids.  The third theme is the computational complexity of auction
\cite{FMS99,FLS99,PU2000,LOS1999,RPH1998}.  For example, Sandholm and
Suri \cite{SS2000} and Akcoglu, Aspnes, DasGupta, and
Kao~\cite{AADKao2000.scp} proposed general frameworks for tackling the
computational hardness of the winner determination problem for
combinatorial auction, which is a special form of multiple-object
auction.

Along these three themes, Kao, Qi, and Tan~\cite{kaoqt.auction.sjp}
considered the {\it position-randomized} multiple-object auction model
specified as follows:
\begin{list}{\thehgt}
{\usecounter{hgt}\setcounter{hgt}{0}\renewcommand{\thehgt}{M\arabic{hgt}}
\setlength{\rightmargin}{0in} 
\settowidth{\leftmargin}{M3} \addtolength{\leftmargin}{\labelsep}}
\item 
There are $m$ bidders competing for $n$ objects, where $m \geq 2$ and $n \geq 1$.
Each bidder has a positive budget and aims to win as many objects as possible.
\item 
Each bidder submits to the auction (1) an initial sequence of $n$ bids
whose total may not exceed the bidder's budget and (2) a randomized
algorithm for permuting the bids. Each bid must be positive or zero.
The final bid sequence that a bidder
actually uses in the auction is obtained by permuting her initial bid
sequence with her bid-permuting algorithm.  The $i$-th bid of each
final sequence is for the $i$-th object.  If an object has $m'$
highest bids, then each of these $m'$ bidders wins this object with
probability $\frac{1}{m'}$.

\item 
Before submitting their initial bid sequences and bid-permuting
algorithms, all bidders know $n$, $m$, and the budget of each
bidder. Furthermore, some bidders may also know the initial bid
sequences and bid-permuting algorithms of others, but not the final
bid sequences.
\end{list}

The assumption M3 addresses the extreme case about informational
security where electronically transmitted information about bids may
be legitimately or illegitimately revealed against the wishes of their
bidders.  To enforce this assumption, the model can be implemented in
an Internet auction as follows.  Before the auction starts, each
bidder submits her initial bid sequence and bid-permuting algorithm to
the trusted auctioneer.  After the auction stops accepting any new
bid, the auctioneer will execute the bid-permuting algorithm publicly.
In such an implementation, while a bidder's initial bid sequence and
bid-permuting algorithm may be leaked to others, her final bid
sequence is not known to anyone including herself and the auctioneer,
until the auction commences.  

Kao et al.~\cite{kaoqt.auction.sjp} also considered an
assumption M3' alternative to M3. Under M3', each bidder may submit
any bidding algorithm which generates a final bid sequence without
necessarily specifying an initial bid sequence. Therefore, less
information may be revealed under M3' than under M3; in other words,
M3' is a weaker security assumption.  Moreover, it is not even clear
that under M3', a bidder's optimal probability distribution of all
possible bids can be computed in finite time.  For these two reasons,
this paper does not use M3'.

Under the above model, Kao et al.~\cite{kaoqt.auction.sjp} gave
optimal bidding algorithms for the case where (1) all bidders have
equal budget, (2) every bid must have a positive dollar amount, and
(3) the number of bidders is two or is an integral divisor of the
number of objects.  In this paper, we resolve only the case of two
bidders where the {\em adversary} bidder $\adv$ knows the {\em
disadvantaged} bidder $\bid$'s initial bid sequence and bid-permuting
algorithm, but not vice versa.  We give a new optimal bidding
algorithm for $\bid$ which improves upon the previous results with two
generalizations: (1) the bidders may have unequal budgets, and (2)
bids with zero dollar amounts are allowed. These two seemingly minor
relaxations make the design and analysis of the new algorithm
considerably more difficult than those of the previous algorithms
\cite{kaoqt.auction.sjp}.  For one thing, one might naturally
anticipate that the optimal expected numbers of objects won by $\adv$
and $\bid$ would be proportional to their budgets.  Surprisingly, this
is not true (Corollary~\ref{cor:limit}).  Our new algorithm runs in
optimal $O(n)$ time in a straightforward manner.  The case with more
than two bidders is open.

To outline the organization of the rest of the paper, we give some
technical definitions first.  The bid set of a bidder refers to the
multiset formed by the bids in her initial bid sequence.  For
convenience, we refer to an initial sequence and its corresponding bid
set interchangeably.  Let $B_\adv$ (respectively, $B_\bid$) be the bid
set of $\adv$ (respectively, $\bid$).  Let $\pi_\adv$ (respectively,
$\pi_\bid$) be the bid-permuting algorithm of $\adv$ (respectively,
$\bid$). $\adv$ may know $\pi_\bid$ and $B_\bid$, while $\bid$ does
not know $\pi_\adv$ and $B_\adv$. We assume that $\adv$ is {\em
oblivious} in the sense that $\adv$ does not know in advance the
outcome of permuting $B_\bid$ with $\pi_\bid$.  Note that bidding
against a non-oblivious adversary is trivial.

Let $w(\pi_\adv,\pi_\bid, B_\adv,B_\bid)$ be the expected number of
objects that $\adv$ wins. Since an auction in our model is a zero-sum
game over the objects, the expected number of objects that $\bid$ wins
is exactly $n-w(\pi_\adv,\pi_\bid,B_\adv,B_\bid)$.  Let
$\winning{\pi_\bid,B_\bid}$ be the maximum of $w(\pi_\adv,\pi_\bid,
B_\adv,B_\bid)$ over all $\pi_\adv$ and $B_\adv$.  We give a bidding
algorithm $(\pi^*_\bid,B^*_\bid)$ which is optimal for $\bid$, i.e.,
\begin{equation}
\label{eq:minmax}
  \winning{\pi^*_\bid,B^*_\bid} = 
  \min_{\pi_\bid,B_\bid}\winning{\pi_\bid, B_\bid}.
\end{equation}
Note that the game has an infinite pure strategy space, so it is not
immediately clear that von~Neumann's min-max theorem is
applicable~\cite{Bla56,BG54,BE1998}.

It has been shown \cite{kaoqt.auction.sjp} that without loss of
generality, (1) $\bid$ always uses the {\it uniform} bid-permuting
algorithm $\unif$ which permutes a sequence $x_1,\ldots,x_n$
with equal probability for every permutation of the indices
$1,\ldots,n$ and (2) thus, $\adv$ uses the {\it identity}
bid-permuting algorithm $\id$ which leaves a sequence
unchanged (see Fact~\ref{fact1a}). Therefore, our main task is to
design an initial bid sequence for $\bid$. A sequence
$x_1,x_2,\ldots,x_\ell$ of bids is {\em proportional} if
$\frac{x_i}{x_j}=\frac{i}{j}$ holds for all $1\leq i,j\leq\ell$. A bid
is {\em unbeatable} if it is greater than the budget of $\adv$. In
this paper, we give a $B^*_\bid$ that consists of (i) a sequence of
zero bids, (ii) a sequence of proportional bids, and (iii) a sequence
of unbeatable bids.  The length of each sequence, which could be zero,
depends on the ratio $\ratio$ of the budget of $\adv$ over that of $\bid$.

Section~\ref{section:algorithm} details
$B^*_\bid$. Section~\ref{section:optimality} proves its optimality for
$\bid$ by showing that Equation~(\ref{eq:minmax}) holds.
Section~\ref{section:conclusion} concludes the paper with open
problems.

\section{The bidding algorithm of the disadvantaged
bidder}\label{section:algorithm} This section gives an optimal bidding
algorithm $(\pi^*_\bid,B^*_\bid)$ for $\bid$. All sets in this paper
are multisets. Let $|X|$ be the number of elements in $X$ counting
multiplicity. Let $X^d=\bigcup_{i=1}^{d} X$, for each positive integer
$d$. Let $X^0=\emptyset$.  Let $\msum{X}=\sum_{x\in X}x$.  Let
$\budget$ be the budget of $\bid$. Hence, the budget of $\adv$ is
$\budget\ratio$.  

We discuss the case $\frac{1}{n}\leq\ratio<n$ first.
Let
$\ourbidset=\setof{0}^{\ell_0}\cup\setof{\frac{\budget}{n-\ell_0}}^{n-\ell_0}$, where
\begin{displaymath}
\ell_0=\left\{
      \begin{array}{ll}
         n & \ratio < \frac{1}{n};\\
         \ceiling{\frac{n-1}{n}}& \ratio=\frac{1}{n};\\
         0 & \ratio > n.
      \end{array}
     \right.
\end{displaymath}
One can easily verify that $(\unif,\ourbidset)$ is an optimal bidding
algorithm for $\bid$, where $\winning{\unif,\ourbidset}$ equals
$\min\setof{\frac{1}{2},\frac{1}{n}}$ for $\ratio=\frac{1}{n}$ and
equals $n-\ell_0$ for $\ratio<\frac{1}{n}$ or $\ratio>n$.

Hence, the rest of the paper assumes $\frac{1}{n}<
\ratio\leq n$. In~\S\ref{section:2.1}, we give a bid set $\ourbidset$
for $\bid$. In~\S\ref{section:2.2}, we prove an upper bound on the
number of objects that $\adv$ can win against
$\ourbidset$. In~\S\ref{section:optimality}, we prove a matching lower
bound, thereby proving the optimality of $\ourbidset$.

\subsection{An optimal bid set for the disadvantaged bidder}
\label{section:2.1}

The next fact simplifies our analysis.
\begin{fact}[See~\cite{kaoqt.auction.sjp}]\label{fact:uniform}
$\text{ }$
\begin{enumerate} 
\item\label{fact1a} If $B_\adv\cap B_\bid=\emptyset$, then
$w(\pi_\adv,\unif,B_\adv,B_\bid)=
w(\id,\unif,B_\adv,B_\bid)\leq
w(\unif,\pi_\bid,B_\adv,B_\bid)$ for any bid-permuting
algorithms $\pi_\adv$ and $\pi_\bid$.

\item\label{fact1b} 
If $\pi_\bid=\unif$, then $\adv$ has an optimal bidding
algorithm with $B_\adv\cap B_\bid=\emptyset$.
\end{enumerate}
\end{fact}
By Fact~\ref{fact:uniform}, the rest of the paper may assume
$\pi_\bid=\unif$ without loss of generality. Thus, let
$\pi^*_\bid=\unif$.  Moreover, as long as $B_\adv$ and
$B_\bid$ are disjoint, we may assume $\pi_\adv=\id$. 

For any positive real numbers $x$ and $y$, define
$\less{x,y}=y\cdot\left(\ceiling{\frac{x}{y}}-1\right)$, 
which is the largest integral multiple of $y$ that is less than
$x$.  Let $\less{x}=\less{x,1}$. Clearly,
$\less{x}=\frac{1}{y}\cdot\less{x y,y}$. Define
\begin{displaymath}
\ourbidset=
\left\{
\begin{array}{ll}
\setof{0}^{\ell_1}\cup
\setof{\frac{\budget}{n-\ell_1}}^{n-\ell_1}&
\text{if $\frac{1}{n}< \ratio\leq\frac{2}{n+1}$};\\
\set{\frac{2i}{\ell_2(\ell_2+1)}\cdot \budget}{i=1,2,\ldots,\ell_2}\cup
\setof{0}^{n-\ell_2}&\text{if $\frac{2}{n+1}< \ratio\leq n$},
\end{array}
\right.
\end{displaymath}
where
\begin{eqnarray*}
\ell_1 &=& \less{2n-\frac{2}{\ratio}+1};\\
\ell_2 &=& \min\setof{n,\floor{\frac{n}{\ratio}}}.
\end{eqnarray*}
Note that $\frac{1}{n}< \ratio\leq \frac{2}{n+1}$ implies
$0<n-\frac{1}{\ratio}<\ell_1< n$. Also, $\frac{2}{n+1}<\ratio\leq n$
implies $1\leq\ell_2\leq n$.  Therefore, $\ourbidset$ is well
defined. Clearly, $\msum{\ourbidset}=\budget$.

\subsection{An upper bound on $\adv$'s winning}
\label{section:2.2}

For each $\ell=1,2,\ldots,n$, let
\begin{eqnarray*}
\ratio_\ell&=&\less{\ratio,\frac{2}{\ell(\ell+1)}};\\
f(\ell)&=&n-\ell+\frac{\ell(\ell+1)\ratio_\ell}{2n}.
\end{eqnarray*}
Define
\begin{displaymath}
{\equilibrium}(n,\ratio)=
\left\{
\begin{array}{ll}
\frac{\ell_1}{n}&\text{if $\frac{1}{n}<\ratio\leq\frac{2}{n+1}$};\\
f(\ell_2)&\text{if $\frac{2}{n+1}<\ratio\leq n$}.
\end{array}
\right.
\end{displaymath}
The next lemma provides an upper bound for
$\winning{\unif,\ourbidset}$.
\begin{lemma}
\label{lemma:lowerbound}
$\winning{\unif,\ourbidset}\leq \equilibrium(n,\ratio)$.
\end{lemma}

\begin{proof}

Case 1: $\frac{1}{n}<\ratio\leq\frac{2}{n+1}$. By
$\ell_1>n-\frac{1}{\ratio}$, we know
$\frac{\budget}{n-\ell_1}>\budget\ratio$. Since $\ourbidset$ contains
$n-\ell_1$ unbeatable bids, the lemma is proved.

Case 2: $\frac{2}{n+1}<\ratio\leq n$.  Let $\ourbidset'$ consist of
the nonzero bids in $\ourbidset$.  It suffices to show that $\adv$
wins no more than $\frac{\ell_2(\ell_2+1)\ratio_{\ell_2}}{2n}$ bids in
$\ourbidset'$ on average.  By Fact~\ref{fact:uniform}(\ref{fact1b}),
$\adv$ has an optimal algorithm $(\id,B_\adv)$ with $B_\adv\cap
\ourbidset'=\emptyset$. Clearly, for each bid $x\in B_\adv$, if $i$ is
the largest index with $\frac{2i\budget}{\ell_2(\ell_2+1)}<x$, then
$x$ wins $\frac{i}{n}$ bids in $\ourbidset'$ on average. Hence, the
unit price for $\adv$ to win a bid in $\ourbidset'$ is greater than
$\frac{2n\budget}{\ell_2(\ell_2+1)}$. By $\pi_\bid=\unif$ and
$B_\adv\cap
\ourbidset'=\emptyset$, the expected number of bids in $\ourbidset'$ that $B_\adv$ 
wins is an integral multiple of $\frac{1}{n}$. 
Since the budget of $\adv$ is $\budget\ratio$, the expected number of bids in $\ourbidset'$ 
that $\adv$ wins is at most
$\less{\frac{\ell_2(\ell_2+1)\budget\ratio}{2n\budget},\frac{1}{n}}=
\frac{1}{n}\cdot\less{\frac{\ell_2(\ell_2+1)\ratio}{2}}=
\frac{\ell_2(\ell_2+1)}{2n}\cdot\less{\ratio,\frac{2}{\ell_2(\ell_2+1)}}=
\frac{\ell_2(\ell_2+1)\ratio_{\ell_2}}{2n}$.
\end{proof}

\section{The optimality of the bid set $\ourbidset$}
\label{section:optimality}
The main result of this section is Theorem~\ref{theorem:main}, which
shows the optimality of $\ourbidset$ by proving
\begin{equation}
\label{eq:optimality}
  \winning{\unif,\ourbidset} = 
  \min_{\pi_\bid,B_\bid}\winning{\pi_\bid, B_\bid}.
\end{equation}
Suppose $B_\bid=\setof{\budget_1,\budget_2,\ldots,\budget_n}$, where $\budget_1\leq \budget_2\leq \cdots
\leq \budget_n$.  Without loss of generality, we may assume
$\msum{B_\bid}=\budget$. Let $B_\ell=\bigcup_{i=1}^{\ell}\setof{\budget_{n-\ell+i}}$
and $t_\ell=\msum{B_\ell}$ for each $\ell=1,2,\ldots,n$. For technical
reason, define $\budget_0=0$.

\subsection{Technical lemmas}
For each $\ell=1,2,\ldots,n$, an {\em $\ell$-set} is a multiset
over $\setof{0,1,\ldots,\ell}$. For any $\ell$-set $I$, let
$\bsum{I,\ell}=\sum_{i\in I}\budget_{n-\ell+i}$. An $\ell$-set $I$ satisfies {\em
Property~P} if the following conditions hold:
\begin{enumerate}[\em{\qquad P}1.]
\item\label{c1} $|I|\leq n$.
\item\label{c2} $\msum{I}\geq \frac{\ratio_\ell\ell(\ell+1)}{2}$.
\item\label{c3} $\bsum{I,\ell}+(n-|I|)\budget_{n-\ell}<\budget\ratio$.
\end{enumerate}
For any positive real number $q$, an $\ell$-set $I$ is an {\em
$(\ell,q)$-set} if $\msum{I}\geq\frac{q\ell(\ell+1)}{2}$ and
$\bsum{I,\ell} \leq q t_\ell$. Clearly, the union of an
$(\ell,q_1)$-set and an $(\ell,q_2)$-set is an
$(\ell,q_1+q_2)$-set. 
\begin{lemma}
\label{lemma:ell-set}
If there is an $\ell$-set set that satisfies Property~P, then
$\winning{\unif,B_\bid}\geq f(\ell)$. 
\end{lemma}
\begin{proof}
Let $I$ be the $\ell$-set that satisfies Property~P.
By Property~P\ref{c1}, the $n$-element set
$X=\setof{\budget_{n-\ell}}^{n-|I|}\cup\set{\budget_{n-\ell+i}}{i\in I}$ is
well defined. By Property~P\ref{c3},
$\msum{X}=\bsum{I,\ell}+(n-|I|)\budget_{n-\ell}< \budget\ratio$. Therefore, there exists
a positive number $\delta$ such that $B_\adv=\bigcup_{x\in
X}\setof{x+\delta}$ satisfies $\msum{B_\adv}\leq \budget\ratio$ and $B_\adv \cap
B_\bid=\emptyset$. Since each bid in $B_\adv$ is greater than
$\budget_{n-\ell}$, $\adv$ wins all $n-\ell$ bids in
$B_\bid-B_\ell$. By Property~P\ref{c2}, the expected number of bids in $B_\ell$
that $\adv$ wins with $B_\adv$ is at least $\frac{\msum{I}}{n}\geq
\frac{\ratio_\ell \ell(\ell+1)}{2n}$. Thus, $\winning{\unif,B_\bid}\geq
n-\ell+\frac{\ratio_\ell\ell(\ell+1)}{2n}=f(\ell)$.
\end{proof}

Roughly speaking, an $(\ell,q)$-set specifies a
good bid set for $\adv$ that spends the budget effectively. For
example, if $I$ is an $(n,\ratio_n)$-set with $|I|\leq n$, then, by $\budget_0=0$
and $\ratio_n<\ratio$, one can easily verify that $I$ satisfies Property~P.  The
next lemma is crucial in designing cost-effective bid sets.

\begin{lemma}\label{lemma:tool}
For each $\ell=1,2,\ldots,n$, the following statements hold.
\begin{enumerate}
\item\label{item:tool1}
For each integer $d\geq 0$, there is an
$\left(\ell,\frac{2d}{\ell}\right)$-set $I_1(\ell,d)$ with $|I_1(\ell,d)|=2d$.

\item\label{item:tool2}
For each integer $h$ with $0\leq h\leq\frac{\ell+1}{2}$, there is an
$\left(\ell,1-\frac{2h}{\ell(\ell+1)}\right)$-set $I_2(\ell,h)$ with $\ell-1\leq
|I_2(\ell,h)|\leq \ell$.

\item\label{item:tool3}
For each integer $k\geq 1$ and each $h=0,1,\ldots,\ell$, there
is an $\left(\ell,k+\frac{2h}{\ell(\ell+1)}\right)$-set $I_3(\ell,k,h)$ with
$k\ell+\floor{\frac{2h}{\ell+1}}\leq|I_3(\ell,k,h)|\leq
k\ell+\ceiling{\frac{2h}{\ell+1}}$.

\item\label{item:tool4}
For each integer $d\geq 1$ and each $h=0,1,\ldots,\ell$, there is an
$\left(\ell,\frac{2d}{\ell}+\frac{2h}{\ell(\ell+1)}\right)$-set $I_4(\ell,d,h)$ with 
$|I_4(\ell,d,h)|\leq 2d+2$.

\item\label{item:tool5}
If $\ell\leq n-1$, then for each integer $d\geq 0$, there is an 
$\ell$-set $I_5(\ell,d)$ with $|I_5(\ell,d)|=2d$, 
$\msum{I_5(\ell,d)}\geq\ell d$, and
$\bsum{I_5(\ell,d),\ell}\leq\frac{2dt_{\ell+1}}{\ell+1}$. 
\end{enumerate}
\end{lemma}
\begin{proof}
Let $L=\setof{1,2,\ldots,\ell}$. For each $i=0,1,\ldots,\ell$, let
$x_i=\budget_{n-\ell+i}$.  Define
$y(i)=\frac{2it_\ell}{\ell(\ell+1)}$, for any integer $i$. Let
$i_0=\arg\max_{i\in L}x_{i}-y(i)$.  Clearly, $x_{i}-x_{i_0}\leq
y(i-i_0)$ holds for each $i\in L$.  By $\sum_{i\in
L}x_i-y(i)=0$, we know $x_{i_0}\geq y(i_0)$.  Let
\begin{eqnarray*} 
i_1&=&\arg\min_{i\in L}x_{i}+x_{\ell-i+1};\\
i_2&=&\arg\max_{i\in L}x_{i}+x_{\ell-i+1};\\
i_3&=&\arg\max_{i\in L-\setof{\ell}}x_{i}+x_{\ell-i}.
\end{eqnarray*}

Statement~\ref{item:tool1}.  Clearly, the inequality
$x_{i_1}+x_{\ell-i_1+1}\leq x_j+x_{\ell-j+1}$ holds for each $j\in
L$. By averaging this inequality over all $\ell$ values of $j$, we
have $x_{i_1}+x_{\ell+1-i_1}\leq \frac{2t_\ell}{\ell}$.  One can
easily verify that the statement holds with
$I_1(\ell,d)=\setof{i_1,\ell-i_1+1}^{d}$.

Statement~\ref{item:tool2}. If $h=0$ (respectively,
$h=\frac{\ell+1}{2}$), then one can easily verify that the statement
holds with $I_2(\ell,h)=L$ (respectively,
$I_2(\ell,h)=I_1\left(\ell,\frac{\ell-1}{2}\right)$).  If $i_0=h$, then
$x_h\geq y(h)$, and thus the statement holds with
$I_2(\ell,h)=L-\setof{h}$.  If $i_0>h$, then $i_0-h\in L$, and thus
the statement holds with
$I_2(\ell,h)=L\cup\setof{i_0-h}-\setof{i_0}$. It remains to prove the
statement for the case $1\leq i_0 < h <\frac{\ell+1}{2}$. Clearly,
$i_0\not\in\setof{\ell-h,\ell+1-h}$ and
$\setof{\ell-h,\ell+1-h,i_0-2h+\ell,i_0-2h+\ell+1}\subseteq L$.  If
$x_{\ell-h}\geq y(\ell-h)$, then the statement holds with
$I_2(\ell,h)=L\cup\setof{i_0-2h+\ell}-\setof{i_0,\ell-h}$.  If
$x_{\ell+1-h}\geq y(\ell+1-h)$, then the statement holds with
$I_2(\ell,h)=L\cup\setof{i_0-2h+\ell+1}-\setof{i_0,\ell+1-h}$. Now we
assume $x_{\ell-h}<y(\ell-h)$ and $x_{\ell+1-h}<y(\ell+1-h)$. If
$\ell$ is even, then clearly $i_2\ne\ell+1-i_2$. One can verify that
the statement holds with
$I_2(\ell,h)=L\cup\setof{\ell+1-h}-\setof{i_2,\ell+1-i_2}$.  If $\ell$
is odd, then clearly $i_3\ne\ell-i_3$. If $x_{i_3}+x_{\ell-i_3}\geq
x_\ell$, then let $J=\setof{i_3,\ell-i_3}$; otherwise, let
$J=\setof{\ell}$. Clearly, $\bsum{J,\ell}\geq
\frac{2t_\ell}{\ell+1}$ and $\msum{J}=\ell$. One can verify that the
statement holds with $I_2(\ell,h)=L\cup\setof{\ell-h}-J$.

Statement~\ref{item:tool3}.  If $h=0$, then the statement holds with
$I_3(\ell,k,h)=L^k$.  If $\frac{\ell+1}{2}\leq h\leq \ell$, then the
statement holds with $I_3(\ell,k,h)=L^{k-1}\cup I_2(\ell,\ell-h)$. If
$x_{h}\leq y(h)$, then the statement holds with
$I_3(\ell,k,h)=L^k\cup\setof{h}$. It remains to consider the case that
both $1\leq h \leq\frac{\ell}{2}$ and $x_{h}>y(h)$ hold.  If
$i_0+2h-\ell-1\in L$ and $i_0\ne h$, then, by $x_h>y(h)$, the
statement holds with $I_3(\ell,k,h)=L^k\cup
I_1(\ell,1)\cup\setof{i_0+2h-\ell-1}-\setof{i_0,h}$. When either
$i_0+2h-\ell-1\not\in L$ or $i_0 = h$ holds, we show $i_0+h\in L$,
which implies that the statement holds with
$I_3(\ell,k,h)=L^k\cup\setof{i_0+h}-\setof{i_0}$.  If $i_0=h$, then
$i_0+h\in L$ holds trivially. If $i_0\ne h$, then, by $2h\leq\ell$, we
know $i_0+2h-\ell-1<i_0$. By $i_0\in L$ and $i_0+2h-\ell-1\not\in L$,
we have $i_0+2h\leq \ell+1$, and thus $i_0+h\in L$.

Statement~\ref{item:tool4}. 
If there is an $i_4\in\setof{0,1,\ldots,h}$ such that
$x_{i_4}+x_{h-i_4}\leq y(h)$, then the statement holds with
$I_4(\ell,d,h)=I_1(\ell,d)\cup\setof{i_4,h-i_4}$.  If there is an
$i_5\in\setof{1,\ldots,\ell-h}$ such that
$x_{h+i_5}+x_{\ell+1-i_5}\leq y(\ell+1+h)$, then, by $d\geq 1$, the statement holds with
$I_4(\ell,d,h)=I_1(\ell,d-1)\cup\setof{h+i_5,\ell+1-i_5}$.  If no such 
$i_4$ or $i_5$ exists, then we have $2t_\ell=\sum_{0\leq
i\leq h}(x_{i}+x_{h-i})+\sum_{1\le
i\leq\ell-h}(x_{h+i}+x_{\ell+1-i})>(h+1)y(h)+(\ell-h)y(\ell+h+1)=2t_\ell$,
a contradiction.

Statement~\ref{item:tool5}. By $\ell+1\leq n$ and
Statement~\ref{item:tool1}, there is an
$\left(\ell+1,\frac{2d}{\ell+1}\right)$-set $I_1(\ell+1,d)$ with
$|I_1(\ell+1,d)|=2d$.  We show that the statement holds with $I_5(\ell,d)=\set{j-1}{j\in
I_1(\ell+1,d)}$. By the proof for Statement~\ref{item:tool1},
$I_1(\ell+1,d)$ is an $(\ell+1)$-set not containing 0. Thus
$I_5(\ell,d)$ is an $\ell$-set. Clearly,
$|I_5(\ell,d)|=|I_1(\ell+1,d)|=2d$,
$\msum{I_5(\ell,d)}=\msum{I_1(\ell+1,d)}-2d\geq(\ell+2)d-2d=\ell d$,
and $\bsum{I_5(\ell,d),\ell}=\bsum{I_1(\ell+1,d),\ell+1}\leq\frac{2d
t_{\ell+1}}{\ell+1}$.
\end{proof}

For each $\ell=1,2,\ldots,n$, let
$\delta_\ell=(\ratio-\ratio_\ell)\frac{\ell(\ell+1)}{2}$. Clearly, 
\begin{equation}
\ratio=\ratio_\ell+\frac{2\delta_\ell}{\ell(\ell+1)}.\label{eq:r0}
\end{equation}
By $\ratio_\ell=\less{\ratio,\frac{2}{\ell(\ell+1)}}$, we know
$0<\delta_\ell\leq 1$.  Let $k_\ell=\floor{\ratio_\ell}$,
$d_\ell=\floor{(\ratio_\ell-k_\ell)\frac{\ell}{2}}$,
$d'_\ell=\floor{(\ratio_\ell-k_\ell)\frac{\ell+1}{2}}$,
$h_\ell=\left(\ratio_\ell-k_\ell-\frac{2d_\ell}{\ell}\right)\frac{\ell(\ell+1)}{2}$,
and
$h'_\ell=\left(\ratio_\ell-k_\ell-\frac{2d'_\ell}{\ell}\right)\frac{\ell(\ell+1)}{2}$.
Since $\ratio_\ell$ is an integral multiple of
$\frac{2}{\ell(\ell+1)}$, we know that $k_\ell$, $d_\ell$, $d'_\ell$,
$h_\ell$, and $h'_\ell$ are integers with $k_\ell=\floor{\ratio_\ell}$,
$0\leq d_\ell<\frac{\ell}{2}$, $0\leq d'_\ell<\frac{\ell+1}{2}$,
$0\leq h_\ell <
\ell+1$, $0\leq h'_\ell< \ell$, and 
\begin{eqnarray}
\ratio_\ell&=&k_\ell+\frac{2d_\ell}{\ell}+\frac{2h_\ell}{\ell(\ell+1)}\label{eq:r1}\\
      &=&k_\ell+\frac{2d'_\ell}{\ell+1}+\frac{2h'_\ell}{\ell(\ell+1)}\label{eq:r2}.
\end{eqnarray}
One can easily verify that either $d'_\ell=d_\ell$ or
$d'_\ell=d_\ell+1$ holds. Moreover, if $d'_\ell=d_\ell$, then
$h'_\ell=d_\ell+h_\ell$. If $d'_\ell=d_\ell+1$, then
$h'_\ell=d_\ell+h_\ell-\ell<\frac{\ell}{2}$. 

\begin{lemma}\label{lemma:k}
For each $\ell=1,2,\ldots,n-1$, we have
\begin{enumerate}
\item\label{item:k1} $k_{\ell+1}=k_\ell$ and 
\item\label{item:k2} $d_{\ell+1}=d'_\ell$.
\end{enumerate}
\end{lemma}
\begin{proof}

Statement~1. Assume for a contradiction that $k_i<k_j$ holds for some
$1\leq i\ne j\leq n$. By $k_j\leq \ratio_j<\ratio$, we know $k_i\leq
k_j-1\leq\ceiling{\ratio}-2$. It suffices to show $\ceiling{\ratio}-k_i\leq 1$
as follows.  If $i$ is even, then, by $d_i<\frac{i}{2}$, we know
$d_i\leq
\frac{i-2}{2}$.  By Equations~(\ref{eq:r0}) and~(\ref{eq:r1}), $\delta_i\leq 1$,
and $h_i<i+1$, we have
$\ceiling{\ratio}-k_i=\ceiling{\frac{2d_i}{i}+\frac{2(h_i+\delta_i)}{i(i+1)}}\leq
\ceiling{\frac{i-2}{i}+\frac{2(i+1)}{i(i+1)}}=1$.
If $i$ is odd, then, by $d'_i<\frac{i+1}{2}$, we know
$d'_i\leq \frac{i-1}{2}$.
By Equations~(\ref{eq:r0}) and~(\ref{eq:r2}), $\delta_i\leq 1$, and $h'_i<i$, we have
$\ceiling{\ratio}-k_i=\ceiling{\frac{2d'_i}{i+1}+\frac{2(h'_i+\delta_i)}{i(i+1)}}\leq 
\ceiling{\frac{i-1}{i+1}+\frac{2i}{i(i+1)}}=1$.

Statement~2. By Equations~(\ref{eq:r0}),~(\ref{eq:r1}),
and~(\ref{eq:r2}) and Statement~1, we have
$\frac{2d_{\ell+1}}{\ell+1}+\frac{2h_{\ell+1}+\delta_{\ell+1}}{(\ell+1)(\ell+2)}
=\frac{2d'_\ell}{\ell+1}+\frac{2(h'_\ell+\delta_\ell)}{\ell(\ell+1)}$.
Therefore,
$d_{\ell+1}+\frac{h_{\ell+1}+\delta_{\ell+1}}{\ell+2}=d'_\ell+\frac{h'_\ell+\delta_\ell}{\ell}$.
By $h'_\ell<\ell$, $h_{\ell+1}<\ell+2$, and
$0<\delta_\ell,\delta_{\ell+1}\leq 1$, we have
$\left|d_{\ell+1}-d'_\ell\right|<1$, and thus $d_{\ell+1}=d'_\ell$.
\end{proof}

\subsection{Matching lower bounds on $\adv$'s winning}
Lemmas~\ref{lemma:main3},~\ref{lemma:main2}, and~\ref{lemma:main1}
analyze the cases (1) $\frac{1}{n}<\ratio\leq\frac{2}{n+1}$, (2)
$\frac{2}{n+1}<\ratio\leq 1$, and (3) $1 < \ratio \leq n$, respectively.
By Lemma~\ref{lemma:k}(\ref{item:k1}), the rest of the section omits the subscript of $k_\ell$.

\begin{lemma}
\label{lemma:main3}
If $\frac{1}{n}<\ratio\leq\frac{2}{n+1}$, then $\winning{\unif,B_\bid}\geq \frac{\ell_1}{n}$.
\end{lemma}
\begin{proof}
Let $\ell$ be the number of bids in $B_\bid$ that are less than
$\budget\ratio$.  If $\ell\geq\ell_1$, then the expected number of
bids that $\adv$ wins with
$B_\adv=\setof{0}^{n-1}\cup\setof{\budget\ratio}$ is at least
$\frac{\ell}{n}$, ensuring $\winning{\unif,B_\bid}\geq\frac{\ell_1}{n}$.
The rest of the proof assumes $\ell<\ell_1$. 
By $(n-\ell)\budget\ratio\leq\sum_{j=\ell+1}^{n}\budget_j$, we have
$\sum_{j=1}^{\ell}\budget_j\leq\budget\ratio\left(\ell-n+\frac{1}{\ratio}\right)$.
By $(n-\ell)\budget\ratio\leq
\budget$, we have $\ell\geq n-\frac{1}{\ratio}$.
By $\ell_1< 2n-\frac{2}{\ratio}+1$, we know $2\ell+1>\ell_1$, which
implies $2\ell\geq\ell_1$.  Let $i^*=\arg\min_{0\leq i\leq
2\ell-\ell_1}\budget_{\ell-i}+\budget_{\ell_1-\ell+i}$.  Let
$X=\setof{0}^{n-2}\cup\setof{\budget_{\ell-i^*},\budget_{\ell_1-\ell+i^*}}$.
Clearly, $\msum X \leq
\frac{2\sum_{j=\ell_1-\ell}^{\ell}\budget_j}{2\ell-\ell_1+1} <
\frac{\sum_{j=1}^{\ell}\budget_j}{\ell-n+\ratio^{-1}} \leq
\budget\ratio$.
 Let  
$B_\adv=\bigcup_{x\in X}\setof{x+\delta}$, where $\delta$ is a number
such that $0<\delta\leq \frac{\budget\ratio-\msum{X}}{n}$ and $B_\adv
\cap B_\bid=\emptyset$. Since $\msum{B_\adv}\leq \budget\ratio$, 
$|B_\adv|=n$, and the expected number of bids that $\adv$ wins with
$B_\adv$ is at least $\frac{\ell_1}{n}$, the lemma is proved.
\end{proof}

\begin{lemma}
\label{lemma:main2}
If $\frac{2}{n+1}<\ratio\leq 1$, then $\winning{\unif,B_\bid}\geq f(\ell_2)$.
\end{lemma}
\begin{proof}
By $\frac{2}{n+1}<\ratio\leq 1$, we know $\ell_2=n\geq 2$ and 
$f(\ell_2)=\frac{(n+1)\ratio_n}{2}$.  By Lemma~\ref{lemma:ell-set}
and $\budget_0=0$, it suffices to show an $(n,\ratio_n)$-set with at
most $n$ elements. If $\ratio_n=\frac{2}{n+1}$, then, by $n\geq 2$, 
$\setof{i^*,n-i^*}$ is a required $(n,\ratio_n)$-set, where
$i^*=\arg\min_{1\leq i\leq n}\budget_i+\budget_{n-i}$. The rest of the
proof assumes $\ratio_n>\frac{2}{n+1}$. Since $\ratio_n$ is an
integral multiple of $\frac{2}{n(n+1)}$, we know
$\ratio_n\geq\frac{2}{n}$.  By $\ratio_n < \ratio\leq 1$ and
Equation~(\ref{eq:r1}), we know
$\ratio_n=\frac{2d_n}{n}+\frac{2h_n}{n(n+1)}$, where $d_n\geq 1$ and $0\leq h_n\leq n$.  By
Lemma~\ref{lemma:tool}(\ref{item:tool4}), we know that $I_4(n,d_n,h_n)$ is
an $(n,\ratio_n)$-set with $|I_4(n,d_n,h_n)|\leq 2d_n+2$. It remains
to consider the case $2d_n+2>n$. By $d_n<\frac{n}{2}$, we have
$d_n=\frac{n-1}{2}$, and thus
$\ratio_n=\frac{n-1}{n}+\frac{2h_n}{n(n+1)}$. By $\ratio_n<1$, we know
$h_n<\frac{n+1}{2}$. It follows that $\ratio_n=1-\frac{2h}{n(n+1)}$,
where $0<h=\frac{n+1}{2}-h_n\leq\frac{n+1}{2}$. By
Lemma~\ref{lemma:tool}(\ref{item:tool2}), $I_2(n,h)$ is an
$(n,\ratio_n)$-set with $|I_2(n,h)|\leq n$.
\end{proof}

\begin{lemma}\label{lemma:main1}
If $1<\ratio\leq n$, then $\winning{\unif,B_\bid}\geq f(\ell_2)$.
\end{lemma}
\begin{proof}
For notational brevity, the proof omits the subscript of $\ell_2$.  By
$1<\ratio\leq n$, we know $\ell=\floor{\frac{n}{\ratio}}$, $1\leq\ell\leq n-1$,
$k\geq 1$, and $\ratio\ell\leq n<\ratio(\ell+1)$.  We first show $f(\ell)\leq
f(\ell+1)$ as follows.  Let $\Delta=f(\ell)-f(\ell+1)$.  Clearly,
$\Delta=1+\frac{1}{n}\left(\ceiling{\frac{\ratio\ell(\ell+1)}{2}}-
\ceiling{\frac{\ratio(\ell+1)(\ell+2)}{2}}\right)$, and thus
$\Delta$ is an integral multiple of $\frac{1}{n}$. Therefore, it
suffices to show
$\Delta<1+\frac{\ratio}{2n}\left(\ell(\ell+1)-(\ell+1)(\ell+2)\right)+\frac{1}{n}=
1-\frac{\ratio(\ell+1)}{n}+\frac{1}{n}<\frac{1}{n}$.

By $f(\ell)\leq f(\ell+1)$ and Lemma~\ref{lemma:ell-set}, it suffices
to show an $\ell$-set or an $(\ell+1)$-set that satisfies Property~P
for each of the following cases.

Case 1: $\ratio\ell\leq n \leq \floor{\ell \ratio_\ell}+k$. Let
$I=I_1(\ell,d_\ell) \cup I_3(\ell,k,h_\ell)$. By
Equation~(\ref{eq:r1}) and Lemmas~\ref{lemma:tool}(\ref{item:tool1})
and~\ref{lemma:tool}(\ref{item:tool3}), we know that $I$ is an
$(\ell,\ratio_\ell)$-set with $\floor{\ell \ratio_\ell}\leq|I|\leq
\ceiling{\ell \ratio_\ell}\leq\ceiling{\ratio\ell}\leq n$, proving
Property~P\ref{c1}. Being an $(\ell,\ratio_\ell)$-set, $I$ satisfies Property~P\ref{c2}
and $\bsum{I,\ell}\leq \ratio_\ell t_\ell$. By $|I|\geq \floor{\ell \ratio_\ell}\geq
n-k$, $k\leq \ratio_\ell<\ratio$, and $\budget_{n-\ell}+t_\ell\leq \budget$, we
know $(n-|I|)\budget_{n-\ell}+\bsum{I,\ell}\leq k \budget_{n-\ell}+ \ratio_\ell
t_\ell<\ratio(t_\ell+\budget_{n-\ell})\leq \budget\ratio$. Therefore, $I$ satisfies
Property~P\ref{c3}.

Case 2: $\floor{\ell \ratio_\ell}+k+1\leq n\leq
k(\ell+1)+2d'_\ell+\floor{\frac{2h'_\ell}{\ell+1}}$. Let
$I=I_5(\ell,d'_\ell)\cup I_3(\ell,k,h'_\ell)$. By
Equation~(\ref{eq:r2}), $k\geq 1$, and $2d'_\ell<\ell+1$, we have
$\floor{\ell \ratio_\ell}+k=
k\ell+\floor{\frac{2d'_\ell\ell+2h'_\ell}{\ell+1}}+k\geq
k\ell+\floor{\frac{2d'_\ell\ell+2h'_\ell+\ell+1}{\ell+1}}\geq
k\ell+2d'_\ell+\floor{\frac{2h'_\ell}{\ell+1}}$. By
Lemmas~\ref{lemma:tool}(\ref{item:tool3})
and~\ref{lemma:tool}(\ref{item:tool5}), we have $n-k\leq
k\ell+2d'_\ell+\floor{\frac{2h'_\ell}{\ell+1}}\leq|I|\leq
k\ell+2d'_\ell+\ceiling{\frac{2h'_\ell}{\ell+1}}\leq\floor{\ell
\ratio_\ell}+k+1\leq n$, proving Property~P\ref{c1}. By
Lemmas~\ref{lemma:tool}(\ref{item:tool3})
and~\ref{lemma:tool}(\ref{item:tool5}) and Equation~(\ref{eq:r2}), we
have $\msum{I}\geq
\ell d'_\ell+\left(k+\frac{2h'_\ell}{\ell(\ell+1)}\right)\frac{\ell(\ell+1)}{2}=
\frac{\ell(\ell+1)}{2}\ratio_\ell$, proving Property~P\ref{c2}. 
By $|I|\geq n-k$, $\budget_{n-\ell}+t_\ell=t_{\ell+1}\leq \budget$, and
Equation~(\ref{eq:r2}), we know $(n-|I|)\budget_{n-\ell}+\bsum{I,\ell}\leq k
\budget_{n-\ell}+\frac{2d'_\ell}{\ell+1}t_{\ell+1}+\left(k+\frac{2h'_\ell}{\ell(\ell+1)}\right)t_\ell
\leq \ratio_\ell \budget<\budget\ratio$, proving Property~P\ref{c3}. 

Case 3: $k(\ell+1)+2d'_\ell+\floor{\frac{2h'_\ell}{\ell+1}}+1\leq
n<\ratio(\ell+1)$.  By $n<\ratio(\ell+1)$, we have $n\leq
\ceiling{\ratio(\ell+1)}-1$. By $\ell+1\leq n$ and
Equations~(\ref{eq:r0}) and~(\ref{eq:r2}), we have
$\ceiling{\ratio(\ell+1)}=k(\ell+1)+2d'_\ell+\ceiling{\frac{2(h'_\ell+\delta_\ell)}{\ell}}$.
By $k(\ell+1)+2d'_\ell+\floor{\frac{2h'_\ell}{\ell+1}}+1\leq
n\leq\ceiling{\ratio(\ell+1)}-1$, we have
$\floor{\frac{2h'_\ell}{\ell+1}}+2\leq\ceiling{\frac{2(h'_\ell+\delta_\ell)}{\ell}}$.
It follows
from $h'_\ell+\delta_\ell\leq \ell$ and $h'_\ell\geq 0$ that 
$\floor{\frac{2h'_\ell}{\ell+1}}=0$ and
$\ceiling{\frac{2(h'_\ell+\delta_\ell)}{\ell}}=2$. 
By $k(\ell+1)+2d'_\ell+\floor{\frac{2h'_\ell}{\ell+1}}+1\leq n\leq
k(\ell+1)+2d'_\ell+\ceiling{\frac{2(h'_\ell+\delta_\ell)}{\ell+1}}-1$,
we know $n=k(\ell+1)+2d'_\ell+1$.  By Lemma~\ref{lemma:k}(\ref{item:k2}) and 
Equations~(\ref{eq:r0}),~(\ref{eq:r1}), and~(\ref{eq:r2}), we have
$\ceiling{\frac{2(h_{\ell+1}+\delta_{\ell+1})}{\ell+2}}=
\ratio(\ell+1)-\left(k(\ell+1)+2d_{\ell+1}\right)=
\ratio(\ell+1)-\left(k(\ell+1)+2d'_{\ell}\right)=
\ceiling{\frac{2(h'_\ell+\delta_\ell)}{\ell}}=2$. Therefore
$1<\frac{2(h_{\ell+1}+\delta_{\ell+1})}{\ell+2}\leq 2$. By
$0<\delta_{\ell+1}\leq 1$ and $\ell\geq 1$, we have
$0<\frac{2h_{\ell+1}}{\ell+2}<2$, and thus
$\floor{\frac{2h_{\ell+1}}{\ell+2}}\leq
1\leq\ceiling{\frac{2h_{\ell+1}}{\ell+2}}$. 
It follows from $n=k(\ell+1)+2d'_\ell+1$, Equation~(\ref{eq:r1}), and
Lemma~\ref{lemma:k}(\ref{item:k2}) that $\floor{\ratio_{\ell+1}(\ell+1)}\leq
n\leq\ceiling{\ratio_{\ell+1}(\ell+1)}$. We prove the statement for the
following two sub-cases.

Case 3(a): $n=\ceiling{\ratio_{\ell+1}(\ell+1)}$. Let
$I=I_1(\ell+1,d_{\ell+1})\cup I_3(\ell+1,k,h_{\ell+1})$. By
Equation~(\ref{eq:r1}), Lemmas~\ref{lemma:tool}(\ref{item:tool1})
and~\ref{lemma:tool}(\ref{item:tool3}), we know that $I$ is an
$(\ell+1,\ratio_{\ell+1})$-set with $\floor{(\ell+1) \ratio_{\ell+1}}\leq|I|\leq
\ceiling{(\ell+1) \ratio_{\ell+1}}=n$, satisfying Property~P\ref{c1}. Being 
an $(\ell+1,\ratio_{\ell+1})$-set, $I$ satisfies Property~P\ref{c2} and
$\bsum{I,\ell+1}\leq \ratio_{\ell+1}t_{\ell+1}$. By $|I|\geq
\floor{(\ell+1) \ratio_{\ell+1}}\geq n-1\geq n-k$, $k\leq
\ratio_{\ell+1}<\ratio$, and $\budget_{n-\ell-1}+t_{\ell+1}\leq \budget$, we know
$\bsum{I,\ell+1}+(n-|I|)\budget_{n-\ell-1}\leq \ratio_{\ell+1} t_{\ell+1}+k
\budget_{n-\ell-1}<\ratio(t_{\ell+1}+\budget_{n-\ell-1})\leq \budget\ratio$, satisfying
Property~P\ref{c3}. 

Case 3(b): $n=\floor{\ratio_{\ell+1}(\ell+1)}$.  Let
$J_1=I_3(\ell,k,0)\cup I_5(\ell,d'_\ell)\cup\setof{h'_\ell}$. Let
$J_2=I_3(\ell,k,0)\cup I_5(\ell,d'_\ell+1)-\setof{h'_\ell}$. By the
proof of Lemma~\ref{lemma:tool}(\ref{item:tool3}), we know
$h'_\ell\in\setof{1,2,\ldots,\ell}\subseteq I_3(\ell,k,0)$. Therefore,
$|J_1|=|J_2|=k\ell+2d'_\ell+1=n-k$. By
$\floor{\frac{2h'_\ell}{\ell+1}}=0$, we know $\ell-h'_\ell\geq
h'_\ell$.  By $\ell-h'_\ell\geq h'_\ell$ and
Lemmas~\ref{lemma:tool}(\ref{item:tool1})
and~\ref{lemma:tool}(\ref{item:tool3}), one can verify that each of
$J_1$ and $J_2$ satisfies Properties~P\ref{c1} and~P\ref{c2}. It
remains to show that either $J_1$ or $J_2$ satisfies
Property~P\ref{c3} as follows. If $\budget_{n-\ell+h'_\ell}<
\frac{2(h'_\ell+\delta_\ell)\budget}{\ell(\ell+1)}$, then, by
$t_{\ell+1}=\budget_{n-\ell}+t_\ell\leq \budget$ and Equations~(\ref{eq:r0})
and~(\ref{eq:r2}), we know $\bsum{J_1,\ell}+(n-|J_1|)\budget_{n-\ell}< 
\frac{2d'_\ell}{\ell+1}t_{\ell+1}+k
t_\ell+\frac{2(h'_\ell+\delta_\ell) \budget}{\ell(\ell+1)}+
k\budget_{n-\ell}
\leq \budget\ratio$. Thus
$J_1$ satisfies Property~P\ref{c3}.  Now we assume
$\budget_{n-\ell+h'_\ell}\geq \frac{2(h'_\ell+\delta_\ell)\budget}{\ell(\ell+1)}$.
By $\ceiling{\frac{2(h'_\ell+\delta_\ell)}{\ell}}=2$, we know
$h'_\ell+\delta_\ell>\frac{\ell}{2}$, and thus
$\ell-(h'_\ell+\delta_\ell)<h'_\ell+\delta_\ell$. It follows from
$t_\ell\leq t_{\ell+1}\leq \budget$ and Equations~(\ref{eq:r0}) and~(\ref{eq:r2}) that
$\bsum{J_2,\ell}+(n-|J_2|)\budget_{n-\ell}<
kt_\ell+\frac{2(d'_\ell+1)t_{\ell+1}}{\ell+1}-\frac{2(h'_\ell+\delta_\ell)\budget}{\ell(\ell+1)}+k\budget_{n-\ell}
\leq \left(k+\frac{2d'_\ell}{\ell+1}+\frac{2(h'_\ell+\delta_\ell)}{\ell(\ell+1)}\right)\budget=\budget\ratio$.
Thus $J_2$ satisfies Property~P\ref{c3}.
\end{proof}

\begin{theorem}\label{theorem:main}
$(\unif,\ourbidset)$ is an optimal bidding algorithm for
$\bid$. Furthermore, $\winning{\unif,\ourbidset}=\equilibrium(n,\ratio)$.
\end{theorem}
\begin{proof}
Clearly, $ \winning{\unif,\ourbidset} \geq
\min_{\pi_\bid,B_\bid}\winning{\pi_\bid, B_\bid}$ holds trivially.
By
Lemmas~\ref{lemma:lowerbound},~\ref{lemma:main3},~\ref{lemma:main2},
and~\ref{lemma:main1}, we know that $\winning{\unif,\ourbidset}\leq
{\equilibrium}(n,\ratio)\leq \winning{\unif,B_\bid}$ holds for any bid set
$B_\bid$ of $\bid$. Therefore, we have Equation (\ref{eq:optimality}),
and thus the equality $\winning{\unif,\ourbidset}=\equilibrium(n,\ratio)$.
\end{proof}

By Theorem~\ref{theorem:main}, the optimal expected winning of $\adv$
(respectively, $\bid$) is $\equilibrium(n,\ratio)$ (respectively,
$n-\equilibrium(n,\ratio)$). We define $\adv$'s {\em effective
winning ratio} $E_\adv(n,\ratio)$ to be
\[\frac{\equilibrium(n,\ratio)}{\frac{n\ratio}{\ratio+1}}.\] 
Similarly, $\bid$'s {\em effective winning ratio} $E_\bid(n,\ratio)$
is \[\frac{n-\equilibrium(n,\ratio)}{\frac{n}{\ratio+1}}.\]

Note that $\frac{\ratio}{\ratio+1}$ (respectively, $\frac{1}{\ratio+1}$)
is the fraction of $\adv$'s (respectively, $\bid's$) budget in the
total budget of $\adv$ and $\bid$.  One might intuitively expect that
$\adv$ (respectively, $\bid$) would win $\frac{n\ratio}{\ratio+1}$
(respectively, $\frac{n}{\ratio+1}$) objects optimally on average.  In
other words, $E_\adv(n,\ratio)=E_\bid(n,\ratio)=1$.
Surprisingly, these equalities are not true, as shown in the next
corollary.

Figures~\ref{fig:plot2} and~\ref{fig:plot1} show $E_\bid(n,\ratio)$ in
3D plots.  Figure~\ref{fig:2d} shows $E_\bid(n,\ratio)$ for some
values of $\ratio$ in 2D plots.

\begin{corollary}\label{cor:limit}$\text{ }$
\begin{enumerate}
\item\label{item:limit1} If $\ratio\geq 1$, then $\lim_{n\rightarrow\infty} E_\adv(n,\ratio)=\frac{(2\ratio-1)(\ratio+1)}{2\ratio^2}$ and $\lim_{n\rightarrow\infty} E_\bid(n,\ratio)=\frac{\ratio+1}{2\ratio}$.

\item\label{item:limit2} If $\ratio\leq 1$, then $\lim_{n\rightarrow\infty} E_\adv(n,\ratio)=\frac{\ratio+1}{2}$ and $\lim_{n\rightarrow\infty} E_\bid(n,\ratio)=\frac{(2-\ratio)(\ratio+1)}{2}$.
\end{enumerate}
\end{corollary}
\begin{proof}
Straightforward.
\end{proof}

{\em Remark.} The formulas in Corollary~\ref{cor:limit} are symmetric
in the sense that those in Statement~\ref{item:limit1} can be obtained
from Statement~\ref{item:limit2} by replacing $\ratio$ with
$\frac{1}{\ratio}$.

\newlength{\heightone}
\setlength{\heightone}{0in}
\addtolength{\heightone}{0.9\textheight}

\begin{figure}
\epsfig{file=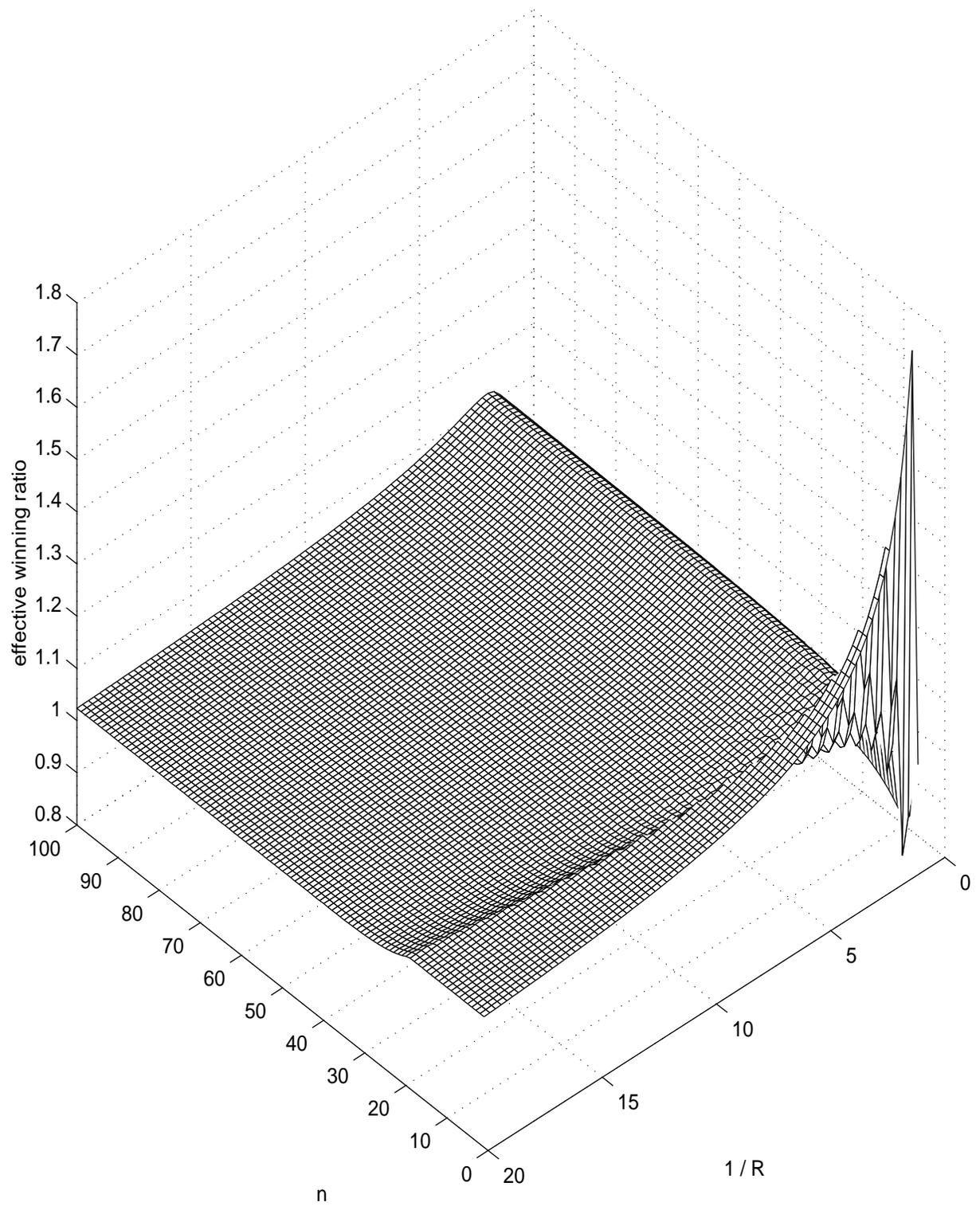,width=\textwidth,height=\heightone}
\caption{The values of $E_\bid$ for $1\leq n\leq 100$ and $\frac{1}{20}\leq \ratio \leq 1$.}
\label{fig:plot2}
\end{figure}

\begin{figure}
\epsfig{file=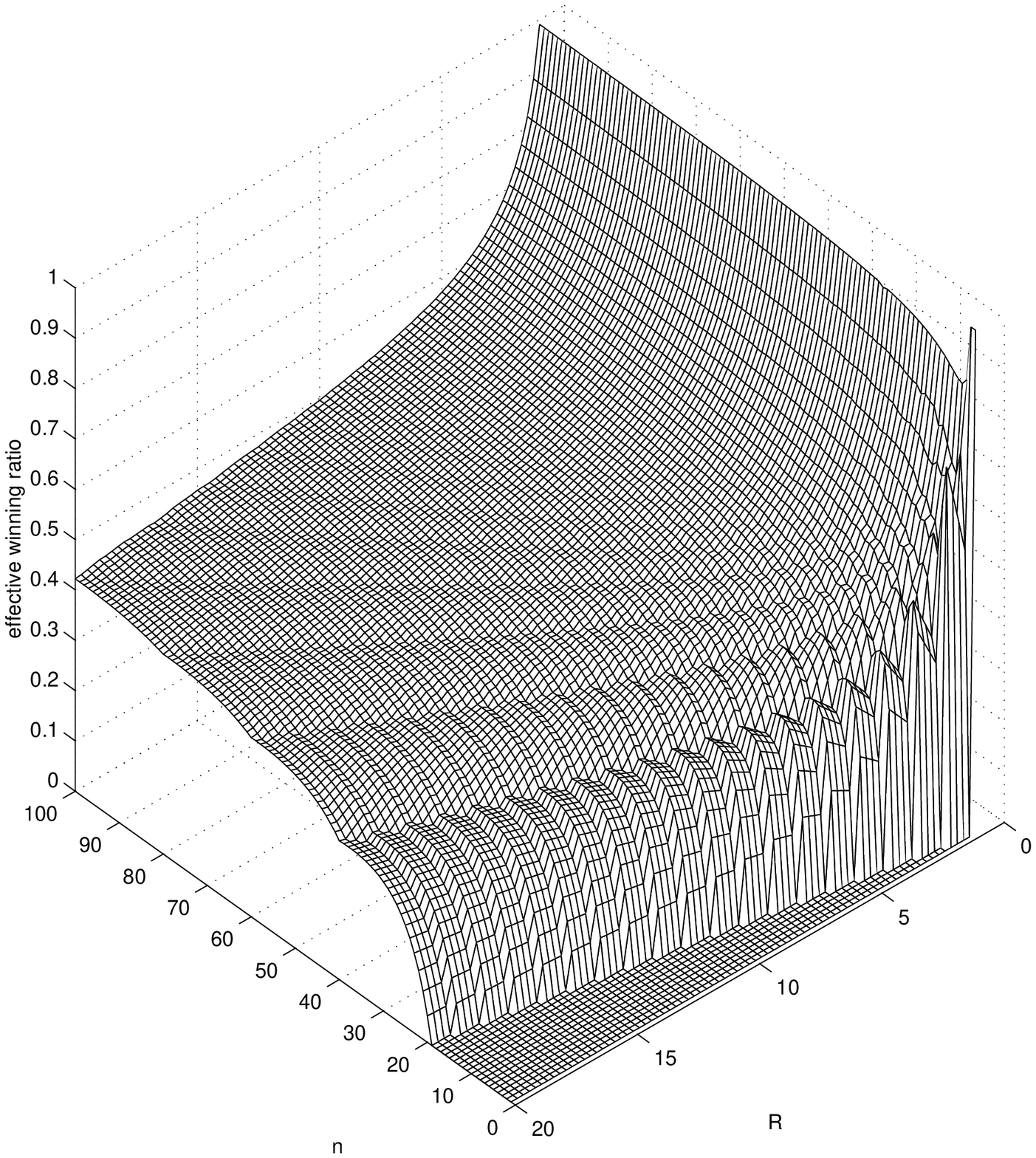,width=\textwidth,height=\heightone}
\caption{The values of $E_\bid$ for $1\leq n\leq 100$ and $1 \leq \ratio \leq 20$.}
\label{fig:plot1}
\end{figure}

\begin{figure}
\epsfig{file=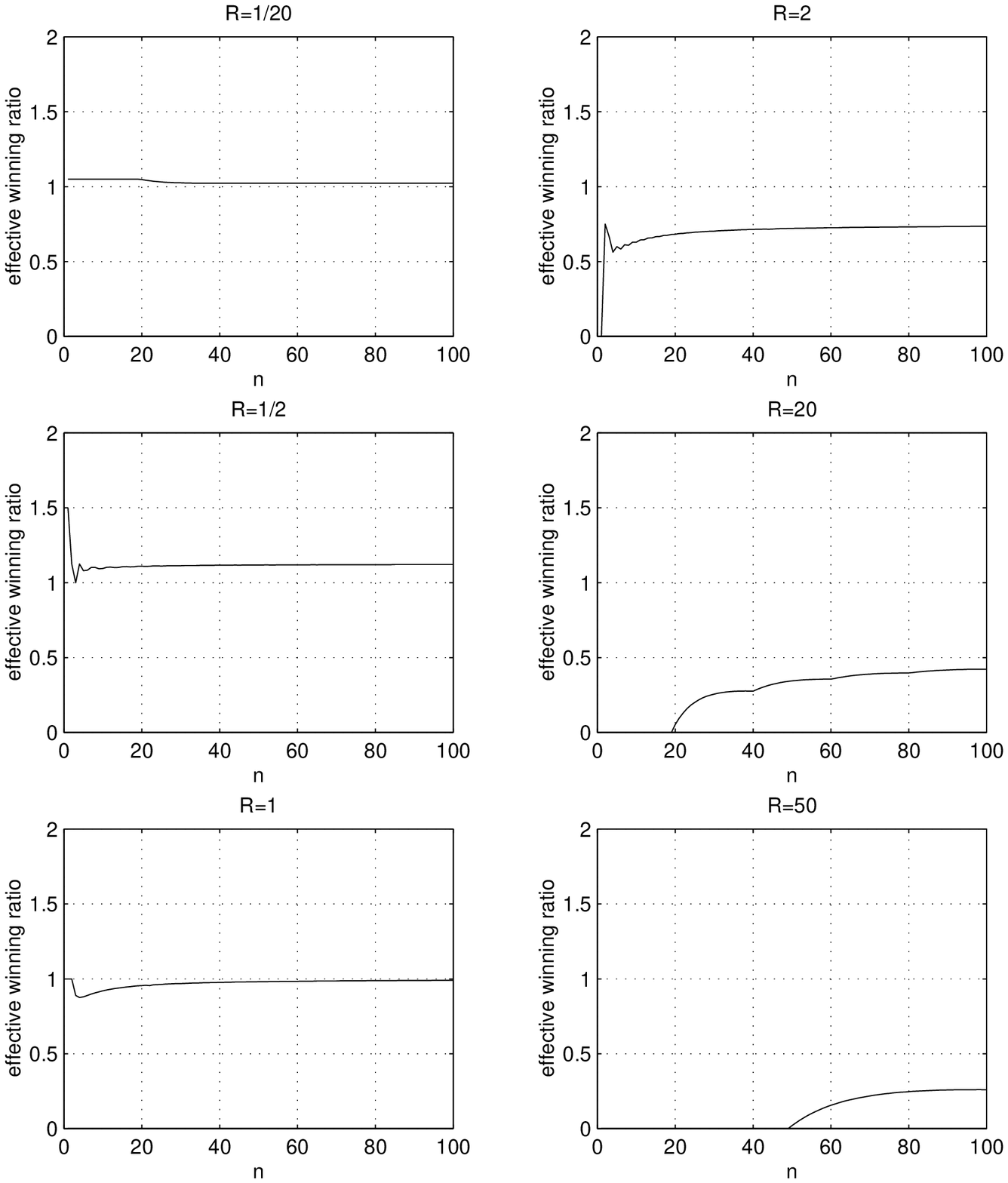,width=\textwidth,height=\heightone}
\caption{The values of $E_\bid$ for $1\leq n\leq 100$ and $\ratio\in\setof{\frac{1}{20},\frac{1}{2},1,2,20,50}$.}
\label{fig:2d}
\end{figure}

\section{Open problems}\label{section:conclusion}
This paper solves the case with two bidders. The case with more than
two bidders remains open.  Another research direction is auction with
{\em collusion}. Note that our model is equivalent to auction with
colluding groups where the bidders all have equal budgets, and those
in the same group pool their money. For example, if the budgets of two
money-pooling bidders are \$100 and \$100, then either of them can
make a bid of \$150.  If pooling is not allowed, then neither can make
a bid of \$150.  It would be of interest to optimally or approximately
achieve game-theoretic equilibria for auctions with non-pooling
collusion.

\bibliographystyle{siam}
\bibliography{all}

\end{document}